\documentclass[12pt]{iopart}
\usepackage{iopams}
\expandafter\let\csname equation*\endcsname\relax
\expandafter\let\csname endequation*\endcsname\relax
\usepackage{amsmath}
\usepackage{graphicx}
\usepackage{caption}
\usepackage{multirow}
\usepackage{xcolor}
\usepackage{color, soul}
\usepackage{float}
 
\usepackage{subcaption}
\usepackage[backend=bibtex,style=numeric,sorting=none]{biblatex}
\begin{document}

\title[]{Single-photon and multi-photon Fano lines for helium and neon using {\tt tRecX-haCC}}

\author{Hareesh Chundayil$^1$, Armin Scrinzi$^2$, Vinay Pramod Majety$^{1*}$}
\address{$^1$ Department of Physics and CAMOST, Indian Institute of Technology Tirupati, Yerpedu, India}
\address{$^2$ Department of Physics, Ludwig Maximilian University, Theresienstrasse 37, Munich, Germany}
\ead{vinay.majety@iittp.ac.in}
\vspace{10pt}

\begin{abstract}
{\color{black}Single-photon and multi-photon ionization of helium and neon atoms by ultrashort extreme ultraviolet radiation is studied. The wavelength of radiation is chosen to excite a set of doubly excited states. We analyse the lineshapes arising from the presence of doubly excited states and numerically demonstrate the sensitivity of the Fano $q$ parameter to the order of multi-photon process. The computations were performed using the {\tt tRecX-haCC} package developed by the authors. The {\tt tRecX-haCC} package solves the time-dependent Schr\"odinger equation in the presence of a laser pulse and computes the photoelectron spectra including the narrow resonance peaks using the tSurff and iSurf techniques. }
\end{abstract}

\section{Introduction}

Interrogation of electron dynamics in the time domain has become a reality with the development of attosecond light sources \cite{Nisoli2017}. These sources typically create a hole in the valence shell and the dynamics can be probed with a second synchronized pulse (extreme ultraviolet (XUV) or infrared (IR)). In this regime, single ionization is the dominant process. At certain photoelectron energies, the ionization probability can exhibit strong modulations due to the presence of doubly excited states. The characteristic signature of these doubly excited states is the asymmetric Fano line shape which arises from the interference of two pathways that lead to the same final ionized state of the system - direct ionization and double excitation followed by decay \cite{Fano1954}. In a seminal work by Ott et al. \cite{ott2013}, it was experimentally shown that the interference could be controlled using an IR pulse and hence the lineshapes. This has generated a lot of interest in better understanding photoionization dynamics in the vicinity of Fano resonances both theoretically and experimentally \cite{Kotur2016,Zielinski2015,Yu2021,Cirelli2018}. {\color{black} New and more intense light sources are being developed at various facilities such as the Extreme Light Infrastructure (ELI) \cite{Shirozhan2024} and FERMI \cite{nandi2024}, which can produce pulses with peak intensities of about 10$^{13}$ W/cm$^2$. With this  multi-photon processes in extreme ultraviolet (XUV) are coming into reach for experimental study. 
}

Accurate description of electron correlation is essential to represent autoionizing states and it is an important benchmark for any new theory that aims to model ultrafast electron dynamics. In this work, we employ our newly developed package, {\tt tRecX-haCC} \cite{Majety_2015,chundayil2024hybrid}, to study multi-photon ionization of atoms - helium and neon. The wavelength of the ionizing XUV pulse is chosen to expose some of the doubly excited states in the system. We compute the photoelectron spectra, extract linewidths and compare with literature. {\color{black}We faithfully reproduce available data for single-photon ionization. In the newly accessible multi-photon regime, an alteration of the q-parameter with the multi-photon order and a tendency to more Lorentzian line shapes at higher orders is observed. }

The properties of doubly excited states are traditionally computed using time independent approaches such as through the diagonalization of the complex scaled electronic Hamiltonian for two electron systems\cite{Jan1997,Lindorth1994} and using R-matrix theory \cite{PHamacher_1989,LIANG2007599}, multichannel quantum defect theory\cite{Nrisimhamurty,Wang} and the time-dependent configuration interaction theory\cite{Carlstrom2022} for many electron systems. While there is a rich literature on one-photon ionization based on these methods, modeling multi-photon ionization to arbitrary photon orders for many electron systems is challenging and is an active area of research \cite{Wang,Benda2021,Andrej2021}. Although the process remains perturbative in nature, the computation of amplitude and phase of the direct transition matrix elements would require the use of elaborate multi-photon perturbation theory. Here, we study the single and multi-photon Fano lineshapes using solutions of the time-dependent Schrödinger equation (TDSE) without making use of perturbation theory. Such a time-dependent approach is the method of choice for short pulses. 

\texttt{tRecX-haCC} solves the TDSE using a hybrid anti-symmetrized coupled channels discretization of the wavefunction. The photoelectron spectra are computed using iSurf\cite{Morales_2016}, which is an infinite time extension to the time-dependent surface flux method (tSurff) \cite{Tao_2012}. In the tSurff method, spectra are computed by integrating the electron flux through a designated surface instead of a direct projection of the final state onto the continuum states. The technique allows us to compute spectra with minimal numerical box sizes. In the iSurf method, the tSurff time-integration is extended from the end of the pulse to infinity, which, in absence of the laser field, can be done in closed form \cite{Morales_2016}. This allows the efficient computation of spectra at low photoelectron energies and the slowly decaying doubly excited states \cite{chundayil2024hybrid,Carlstrom2022}.

The article is organized as follows: In section \ref{sec-th}, we briefly describe the tRecX-haCC method followed by a description of the tSurff and iSurf methods in section \ref{sec-surf}. Finally, we present our results on multi-photon ionization of helium and neon in section \ref{sec-res}.

\section{tRecX-haCC method} \label{sec-th}

Consider a $N$ electron system in the Coulomb field of a nucleus of charge $Z$. Starting from the ground state of the multielectron
Hamiltonian (in a.u.),
\begin{equation} \label{eq-ham}
\hat{H}_{0}=\sum_{i=1}^{N}\left[-\frac{\nabla_{i}^{2}}{2}-\frac{Z}{r_i}\right]+\sum_{i=1}^{N}\sum_{j=1}^{i-1}\frac{1}{|\vec{r}_{i}-\vec{r}_{j}|}    
\end{equation}
the time evolution of the state in the presence of an external light field is computed by solving the time-dependent Schr\"odinger equation 
\begin{equation}
i\frac{\partial}{\partial t}|\Psi\rangle=(\hat{H}_{0}+\hat{H}_{1})|\Psi\rangle.    
\end{equation}
Here, $\hat{H}_{1}$ represents the light-matter interaction operator
in the mixed gauge. {\color{black}In Ref. \cite{mixed2015}, it was shown that a certain combination of the length and modified velocity gauge is an optimum choice when solving the TDSE using hybrid basis sets such as ours. See also discussion in \cite{chundayil2024hybrid}.} The mixed gauge operator in the dipole approximation can be computed starting from the length form given by
\begin{equation}
\hat{H}_{L}=-\sum_{j=1}^{N}\vec{E}\cdot\vec{r}_{j}
\end{equation}
with $\vec{E}(t)$ being the electric field and applying a gauge transform
\begin{equation}
\hat{U}_{g}=\prod\limits_{j=1}^{N} \hat{U}_{g}^{(j)}
\end{equation}
with 
\begin{equation} \label{eq-gauge}
\hat{U}_{g}^{(j)}=\begin{cases}
1 & |\vec{r}_{j}|\leq r_{g} \\
e^{i\vec{A}(t)\cdot\vec{r}_j(1-\frac{r_g}{|\vec{r}_j|})} & |\vec{r}_{j}|>r_g
\end{cases}
\end{equation}
and $\vec{A}(t)=-\int_{-\infty}^{t}\vec{E}(t')dt'$ being the vector potential.

The wavefunction is discretized using a hybrid coupled channels basis
composed of 
\begin{enumerate}
    \item Neutral states ($|\mathcal{N}\rangle$) which are approximate eigenstates of the $N$ electron system. These are computed using configuration interaction (CI) theory. The CI states are linear combinations of Slater determinants constructed using Gaussian based atomic/molecular orbitals resulting from a Hartree-Fock calculation. We used the \texttt{COLUMBUS} package \cite{Lischka2011} for this purpose. 
 \item Channel functions which are anti-symmetrized 
products of ($N-1$) electronic ionic states ($|I\rangle$) and a
numerical one-electron basis. The ionic states are again computed
using \texttt{COLUMBUS}. The numerical one electron basis
is composed of two parts
  \begin{enumerate}
      \item Gaussian based atomic/molecular orbital basis ($|i\rangle$)
\item A general single centered expansion ($|\alpha\rangle$) constructed using finite element discrete variable representation (FE-DVR) basis on the radial coordinate and spherical harmonics on the angular coordinates. In addition, these are made orthogonal to the molecular orbital basis as $|\alpha_{\perp}\rangle$ = $|\alpha\rangle-\sum_i |i\rangle \langle i|\alpha\rangle$.
  \end{enumerate}
\end{enumerate}

The hybrid anti-symmetrized coupled channels (haCC) expansion for the wavefunction is written as
\begin{equation} \label{eq-bas}
|\Psi\rangle=\underbrace{\sum_{\mathcal{N}}C_{\mathcal{N}}(t)|\mathcal{N}\rangle}_{\text{Neutrals}}+\underbrace{\sum_{I,i}C_{I,i}(t)\mathcal{A}|I,i\rangle+\sum_{I\alpha}C_{I,\alpha}(t)\mathcal{A}|I,\alpha_{\perp}\rangle}_{\text{Channel functions}}.
\end{equation}

Here, $\mathcal{A}$ denotes the antisymmetrizer and $C_{\mathcal{N}},$$C_{I,i}$ and $C_{I,\alpha}$ are the time-dependent expansion coefficients that represent the dynamics. The expansion captures essential parts of the Hilbert space required for single ionization problems. The basis respects exchange symmetry, includes correlation in the initial state that somewhat exceeds the original \texttt{COLUMBUS} description due to the additional channel functions, and accounts for the "dynamical correlation", i.e. correlation that builds up during the excitation and ionization process.
We label various haCC calculations using the notation haCC(n,i) where the n and i energetically lowest neutral and ionic states, respectively, are used, unless indicated otherwise. 

Approximating the Hilbert space by a large finite dimensional space $ \{|\mathcal{N}\rangle\} \oplus \{\mathcal{A}|I,i\rangle\} \oplus  \{\mathcal{A}|I,\alpha_{\perp}\rangle\}$ converts the time-dependent Schr\"odinger equation into a set of coupled ordinary differential equations for the time-dependent coefficients which are then solved using adaptive Runge-Kutta methods. Absorbing boundary conditions are imposed using the infinite range exterior complex scaling method \cite{irECS}. This results in non-hermitian matrix representation of operators.

The hybrid anti-symmetrized coupled channels (haCC) basis (\ref{eq-bas}) is non-orthogonal and different operators evaluated in our basis have different degrees of sparsity. These aspects are efficiently handled using flexible tree data structures that are part of the {\tt tRecX} package \cite{SCRINZI2022108146}. We refer the readers to \cite{SCRINZI2022108146,chundayil2024hybrid} for further details of implementation of the solver. In the next section, we will describe the spectral analysis method.

\section{Spectral analysis using tSurff and iSurf} \label{sec-surf}

{\color{black}The tSurff method was first proposed in Ref.~\cite{Tao_2012, Scrinzi2012} and it was extended in Ref.~\cite{Morales_2016} to include iSurf for the long-time emission after the passage of the pulse. Here we briefly summarize the methods and discuss minor points that are specific to haCC.}

Consider a photoionization process that ejects a photoelectron with
momentum $\vec{k}$ and leaves the residual ion in state $|I\rangle$. If the final state is $|X_{I,\vec{k}}\rangle$, the probability density for the photoionization to final momentum $\vec{k}$ can be computed as 
\begin{equation} \label{eq-pes}
P_{I,\vec{k}}=|\underbrace{\langle X_{I,\vec{k}}|\Psi(T)\rangle}_{b_{I,\vec{k}}}|^{2}
\end{equation}
where $T>T_p$ is any time after the end of the laser pulse at $T_p$. 

{\color{black} The computation of $b_{I,\vec{k}}$ by Eq.~(\ref{eq-pes}) is complicated by the fact that with longer pulses $\Psi(T_p)$ may be very large in configuration space. For example, at the near IR wavelength of 800 nm, the wavefunctions can grow to 100's of atomic units during a single optical cycle. Further, one needs to determine the channel continuum function $|X_{I,\vec{k}}\rangle$, which involves the repeated solution of large linear systems. The tSurff method avoids both problems by recording the flux as it leaves the interaction volume and accounting for the later modification of momenta by the dipole laser field.}

{\color{black} In tSurff one assumes that for electrons outside a certain radius $R_c$ all interactions with the remaining electrons and with the nuclei can be neglected, and in that region the final state assumes the asymptotic form}
\begin{equation}
|X_{I,\vec{k}}\rangle \sim \mathcal{A}\left(|I\rangle\otimes|\vec{k}\rangle\right)=:|I,\vec{k}\rangle
\end{equation}
with a plane wave for $|\vec{k}\rangle$. 
Picking some large time $T$, where the unbound electron with coordinate $\vec{r}_n$ has moved beyond the radius $R_c$, we approximate the $b_{I,\vec{k}}$ in equation (\ref{eq-pes}) as

\begin{equation}\label{eq-bik}
\langle X_{I,\vec{k}}|\Psi(T)\rangle \approx b_{I,\vec{k}}(T)=\sum_{n=1}^N \langle I,\vec{k}|\Theta(r_n-R_c)|\Psi(T) \rangle,
\end{equation}
where $\Theta(r_n-R_{c})$ is the  Heaviside function, which is $1$ if the
argument is positive and 0 otherwise. 

As described in Refs. \cite{Majety_2015,Tao_2012}, the volume integral in equation (\ref{eq-bik}) can be converted to a surface and a time integral, which is computationally more tractable. 
For times before the end of the laser pulse, $t<T_p$, the action of the laser field on the emitted electron needs to be taken into account. The time-evolution of the remote electron is described by the time-dependent Volkov state
\begin{equation} 
|\vec{k},t\rangle=\exp \left[-i\int_0^t \frac{k^2}{2}-\vec{k}\cdot\vec{A}(\tau) d\tau\right]|\vec{k},0\rangle
\end{equation}
which solves the time-dependent Schr\"odinger equation of the free electron in velocity gauge
\begin{equation}\label{eq-volkov}
i\frac{d}{dt}|\vec{k},t\rangle = \left[-\frac12 \Delta + i\vec{\nabla}\cdot\vec{A}(t)\right]|\vec{k},t\rangle.
\end{equation}
In the multi-electron case \cite{Scrinzi2012} one also needs to include the time-evolution of the ion due to laser-coupling between the ionic bound states as
\begin{equation}\label{eq-ion}
i\frac{d}{dt}|I,t\rangle = \sum_{J}H^{\rm (ion)}_{IJ}(t) |J,t\rangle,
\end{equation}
where $H^{\rm (ion)}_{IJ}(t)$ is the time-dependent Hamiltonian for the ionic system and the sum is over all ionic bound states included in the haCC expansion.
The ionic Schr\"odinger equation (\ref{eq-ion}) generates a unitary time evolution, which relates the ionic state $|I,t\rangle$ at time $t$ to the states at time $t=0$ by
\begin{equation}
    |I,t\rangle = \sum_J U^{\rm (ion)}_{IJ}(t) |J,0\rangle.
\end{equation}
Taking the time-derivative of equation (\ref{eq-bik}) and integrating over time, one obtains the 
\begin{equation} \label{eq-tsurff}
b_{I,\vec{k}}(T)=N \sum_J U^{\rm (ion)\,\dagger}_{IJ}(T)\int_{0}^{T}dt\langle J,\vec{k},t|\hat{C}|\Psi(t)\rangle
\end{equation}
with the notation $|I,\vec{k},t\rangle:=\mathcal{A}(|I, t\rangle\otimes |\vec{k},t\rangle)$ and the commutator
\begin{equation}
    \hat{C} = [-\frac{1}{2}\nabla_N^{2}+i\vec{A}(t)\cdot\vec{\nabla}_N,\Theta(r_N-R_{c})].
\end{equation}
The commutator of the derivatives with the $\Theta$-functions in $\hat{C}$ consists of terms that contain the $\delta$-function at $r_n=R_c$ and its derivative.
This reduces the evaluation of the matrix element to an integral over the surface of the sphere with radius $R_c$ involving values and derivatives of the wavefunctions on the surface. In equation (\ref{eq-tsurff}) the factor $N$ arises from the summation over all electrons and the inverse of the ionic time-evolution $U^{\rm (ion)\,\dagger}(T)$  accounts for the evolution of different $|J,t\rangle$ states  towards the final $|I,T\rangle$ state. Also, note that equation (\ref{eq-volkov}) is in velocity gauge. By our use of mixed gauge, surface values and derivatives are in length gauge up to $r_g$, see equation (\ref{eq-gauge}). Before applying equation (\ref{eq-tsurff}) wavefunction values and derivatives need to be  converted to velocity gauge. 
We have excluded multiple ionization from our discussion. The extension of tSurff to multiple ionization is discussed in Ref.~\cite{Scrinzi2012}, but that has not been implemented for haCC so far. 

To account for all photoelectrons including near zero energy photoelectrons and those due to delayed emission as in a slowly decaying resonance, one needs to integrate up to very large times long after the end of the pulse, $T\gg T_p$. 
{\color{black}
In this case it can be efficient to supplement the spectral amplitude collected at $R_c$ up to time $T_p$ with a spectral analysis of the wavefunction $|\Psi(T_p)\rangle$ at $|\vec{r}_i|<R_c$ \cite{Morales_2016}. This is easily achieved in the spectral representation of $\hat{H}_0$, where one can compute the time-integral after $T_p$ analytically.
We write the spectral decomposition of $\hat{H}_0$ as 
}

\begin{equation}
\hat{H}_{0}=\sum_s |s\rangle E_{s}\langle \tilde{s}|.
\end{equation}where $|s\rangle$ and $\langle \tilde{s}|$ are the right and left eigenvectors of $\hat{H}_0$. As we use  absorbing boundary conditions,
the $E_s$ are complex in general. The left and right eigenvectors still form a resolution of identity $1=\sum_s |s\rangle\langle \tilde{s}|$, but they
are not trivially related to each other by complex conjugation, see Ref.~\cite{irECS}.
The state at the end of the pulse represented in the haCC basis $|h\rangle$ can be re-expressed in the spectral basis as
\begin{equation}
|\Psi(T_p)\rangle=\sum_{h}|h\rangle c_{h}(T_p)=\sum_{hs}|s\rangle \langle \tilde{s}|h\rangle c_{h}(T_p)=:\sum_{s}|s\rangle d_{s}(T_p).
\end{equation}
After the end of the pulse $t>T_p$, the time evolution of  $|\Psi(t)\rangle$ is
\begin{equation} \label{eq:psiEv}
    |\Psi(t)\rangle=\sum_{s}e^{-iE_{s}(t-T_p)}|s\rangle d_{s}(T_p).
\end{equation}
and the continuum state $|I,\vec{k},t\rangle$ 
evolves as
\begin{equation} \label{eq:Xev}
|I,\vec{k},t\rangle=|I,\vec{k},T_p\rangle e^{-i(E_{I}+\frac{1}{2}k^{2})(t-T_p)},    
\end{equation}
where $E_I$ is the energy of the ion and $\frac{1}{2}k^2$ the energy of the free electron.  

Using equations (\ref{eq:psiEv}), (\ref{eq:Xev}), the time integral in (\ref{eq-tsurff}) can be extended to infinity as:

\begin{align}
b_{I,\vec{k}}(t & =\infty)=b_{I,\vec{k}}(T_p)+\int_{T_p}^{\infty}dt \langle I,\vec{k},t|\hat{C}|\Psi(t)\rangle \nonumber\\
 & =b_{I,\vec{k}}(T_p)-i\sum_{s}\langle I,\vec{k},T_p|\hat{C}|s\rangle\frac{1}{E_{s}-E_{I}-\frac{1}{2}k^{2}} d_{s}(T_p)
 \label{eq-isurf1}
\end{align}
where $b_{I,\vec{k}}(T_p)$ is computed using (\ref{eq-tsurff}). As the $E_s$ are complex the inverses are well-defined.
Note that in absence of the laser field there is no coupling between ionic channels and hence no summation over ionic channels appears in the second term.
Equation (\ref{eq-isurf1}) can be written as 
\begin{equation} \label{eq-isurf2}
b_{I,\vec{k}}(t=\infty)=b_{I,\vec{k}}(T_p)-i\langle I,\vec{k}.T_p|\hat{C}(\hat{H}_0-E_{I}-\frac{1}{2}k^{2})^{-1}|\Psi(T_p)\rangle.
\end{equation}
When the spectral decomposition is not available, iterative methods can be employed to obtain $(\hat{H}_0-E_{I}-\frac{1}{2}k^{2})^{-1}|\Psi(T_p)\rangle$.
Denoting the photoelectron energy as $E$$(=\frac{k^2}{2})$, the photoionization yield corresponding to emission into channel $I$ is computed by integrating over directions as
\[
Y_{I}(E) = \int k^2 |b_{I,\vec{k}}(t=\infty)|^2 d\Omega_k  
\]and the total yield for a photoelectron energy is
\[
Y(E) = \sum_I Y_I(E)
\]

\section{Results} \label{sec-res}

We study photoionization of helium and neon atoms by few cycle linearly polarized laser pulses. As defined in Ref. \cite{SCRINZI2022108146}, the vector potential has the form
\begin{equation}
\vec{A}(t) = \vec{\alpha}(t) \sqrt{\frac{I_0}{2 \omega^2}} \sin(\omega t - \phi)
\end{equation}
where $I_0$ is the peak intensity of the pulse, $\omega$ is the central frequency, $\phi$ is the carrier envelope phase and $\vec{\alpha}(t)$ defines the polarization and the pulse envelope. The pulse duration is defined in terms of its full width at half maximum (FWHM) and in the units of the number of optical cycles at the central wavelength \cite{SCRINZI2022108146}. We choose a $\cos^8$ pulse shape rather than the $\cos^2$ shape often used for calculations. This is to avoid any numerical artifacts that can appear with a $\cos^2$ pulse but are absent with $\cos^8$ pulse, see Ref. \cite{Zielinskidouble} for further discussion. In fact, keeping FWHM fixed, $\cos^N$ converges towards a Gaussian as $N\to\infty$, which further motivates the use of the higher power of $N=8$.

The wavelengths are chosen to expose doubly excited states. At each chosen wavelength, we present total and channel resolved photoelectron spectra and analyze select lineshapes resulting from the presence of doubly excited states. We parameterize the lineshapes using the equation
\begin{equation} \label{eq-Fano}
    Y_{I}(E)=Y_{a}\frac{(q+\epsilon)^{2}}{1+\epsilon^{2}}+Y_{b}
\end{equation}
Here, $\epsilon=\frac{E-E_{r}}{\frac{1}{2}\Gamma}$, $E_{r}$ is the position of the resonance, $\Gamma$ is the width, $q$ is the Fano parameter which characterizes the line profile, $Y_{a}$ and $Y_{b}$ are slowly varying background yields \cite{Fano1965}. 
The original Fano parametrization was given for photoionization cross-sections, but we employ the same for the yields. This is justified as our few cycle pulses have a broad bandwidth and the spectral intensity of the laser pulse can be assumed to remain constant over the width of a given resonance. Parameters $Y_a$ and $Y_b$ can be used to construct the correlation coefficient $\rho^2=\frac{Y_a}{Y_a+Y_b}$ which is equal to 1, if the ionization process involves a single channel \cite{Fano1965}. Where multiple channels are involved, the correlation coefficient deviates from 1 due to a background from the other channels.

\subsection{Helium}

We present here total and channel resolved photoelectron spectra for helium ionized by 2-cycle XUV pulses with central wavelengths of 20 nm($\lambda_1$), 40 nm($\lambda_2$), 60 nm($\lambda_3$) and 80 nm($\lambda_4$) and a peak intensity of $10^{15}$ W/cm$^2$. The wavelengths were chosen such that a set of doubly excited states can be reached by absorption of $n=1,2,3 \text{ and } 4$ photons. The quantum chemical calculations to construct the ionic and neutral states were performed using the aug-cc-pvtz basis set. A haCC(1,9) basis set was then constructed and photoelectron spectra were computed. The ground state energy with the haCC(1,9) basis is -2.901 a.u. The $1s$ channel energy is -1.999 a.u.. The ionization potential deviates from the exact value by about 30 meV.

\begin{figure}[H]
    \centering
    \includegraphics[scale=0.35]{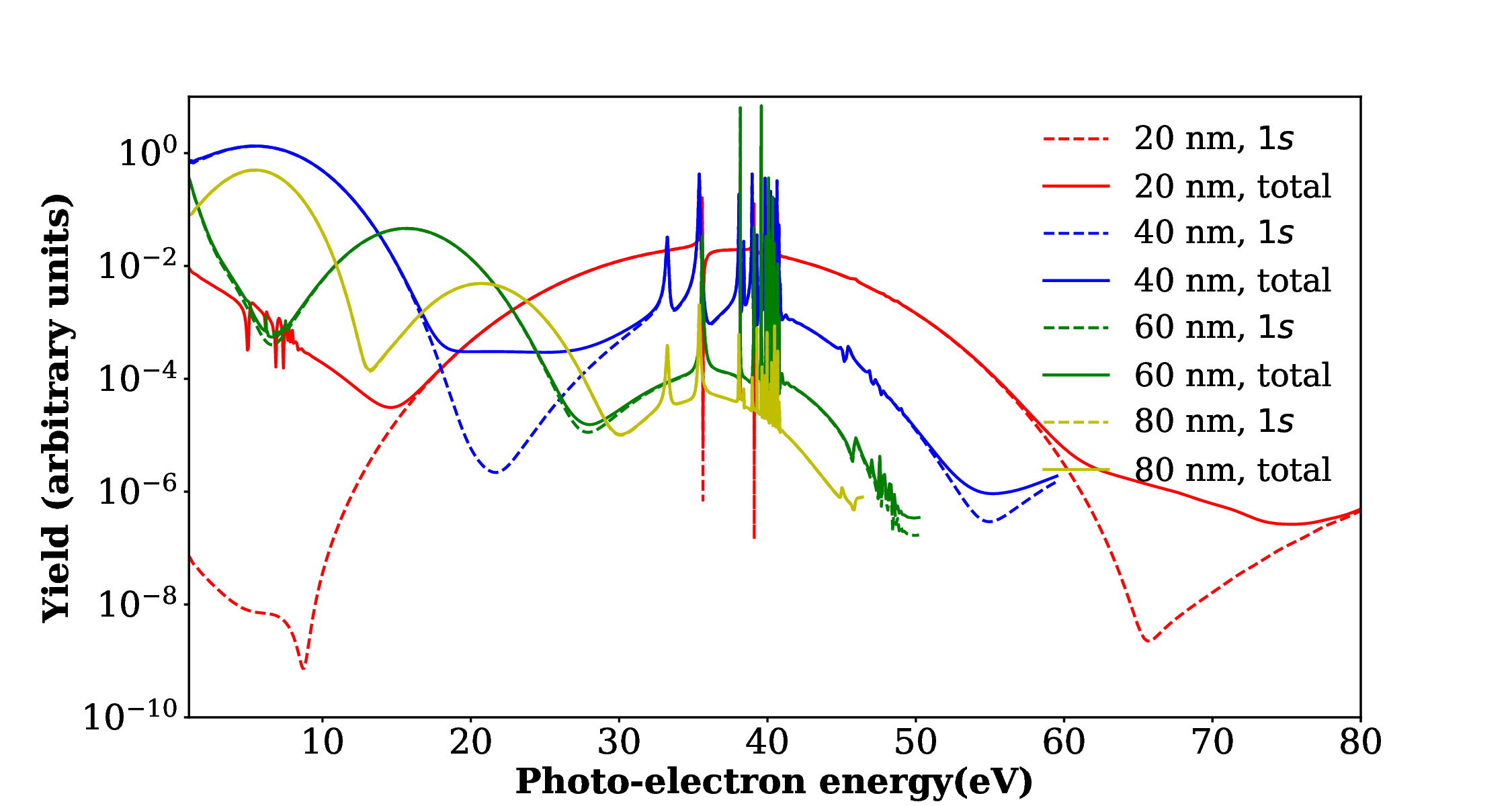}        
    \caption{Total (solid) and $1s$ channel (dashed) photoelectron spectra for helium computed using haCC(1,9) basis with 2-cycle laser pulses having central wavelengths of 20 nm, 40 nm, 60 nm and 80 nm.}
    \label{Fig:He-total}
\end{figure}

\begin{figure}[H]
    \centering
    \includegraphics[scale=0.3]{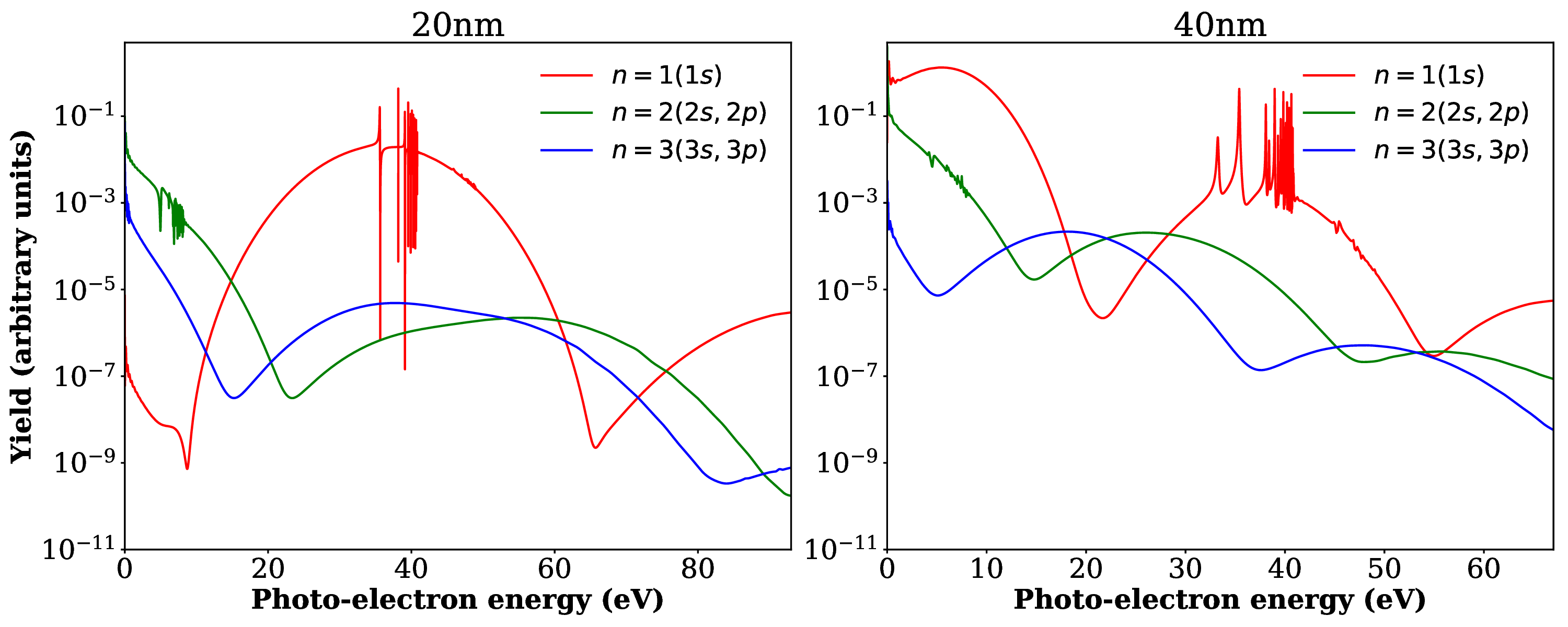}
    \caption{Channel resolved spectra for helium computed using haCC(1,9) basis with 2-cycle laser pulses having central wavelengths of 20 nm and 40 nm. The channel label is indicated in the legend.}
    \label{Fig:He-channel20}
\end{figure}

\begin{figure}[H]
    \centering
    \includegraphics[scale=0.3]{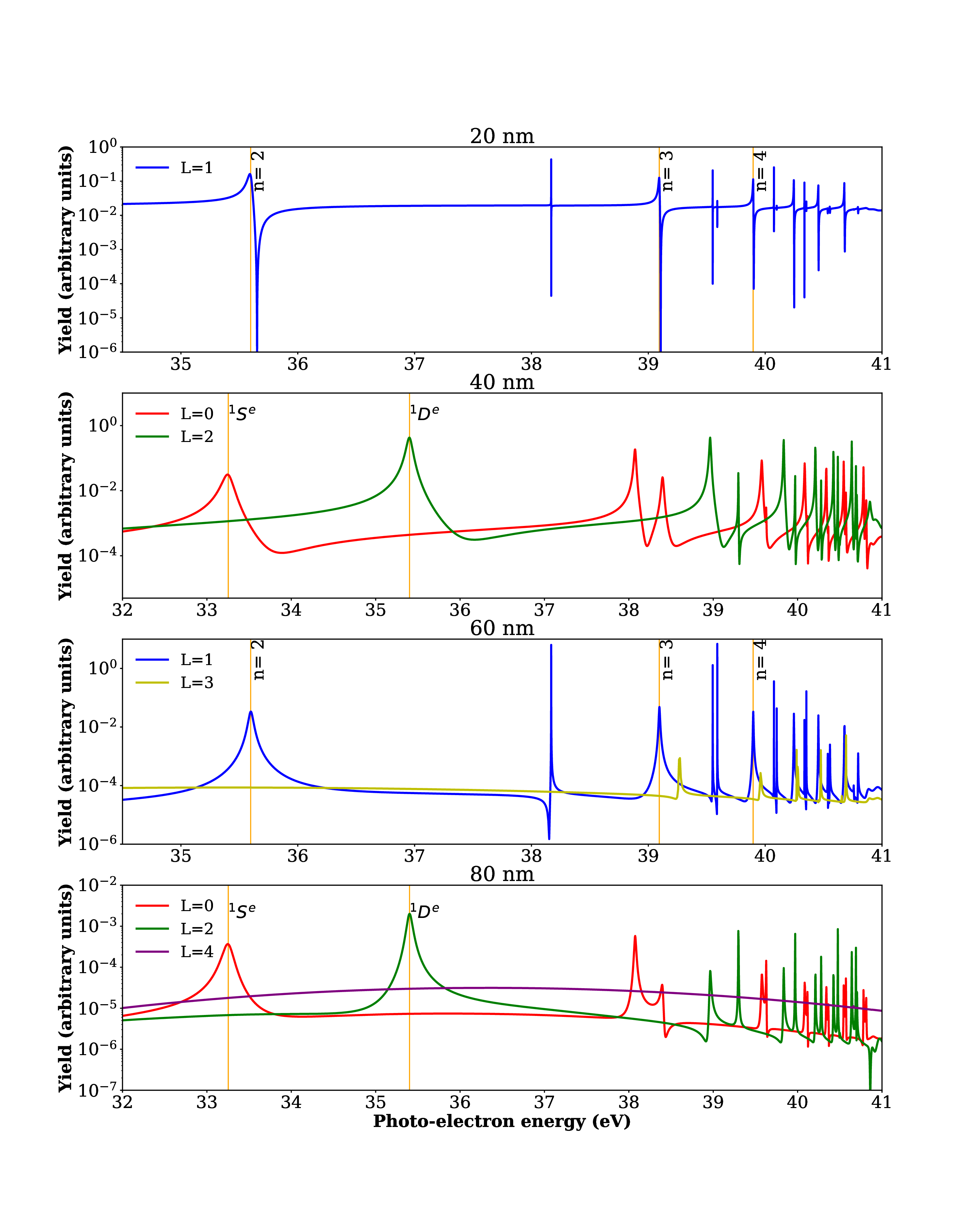}
    \caption{Partial wave analysis of $1s$ channel spectra for helium in the photoelectron energy window 32-42 eV. Each sub-plot title indicates the excitation wavelength. Dominant partial waves are presented.}
    \label{Fig:He1s-zoomed}
\end{figure}

Figure \ref{Fig:He-total} presents the total and $1s$ channel photoelectron spectra at all the considered wavelengths. We observe multi-photon absorption peaks at positions defined by $n\hbar\omega-I_p$ where $I_p$ is the ionization potential. The width of these peaks is related to the bandwidth of the laser pulse. {\color{black} 
While major contribution to photoionization comes from the $1s$ channel, at 20nm and 40 nm contributions from the n=2 and n=3 channels are important at certain photoelectron energies. These are the regions where ionization to $1s$ channel is not energetically favourable. Channel resolved spectra demonstrating this are presented in figure~\ref{Fig:He-channel20}. At 60nm and 80nm, the $1s$ channel is dominant in the entire range of photoelectron energies considered and hence the total and the $1s$ channel spectra coincide.} 
We also observe several rapid modulations around 10 eV, 32-42 eV and 45-50 eV. These features result from the presence of doubly excited states. In this work, we analyze the features in the photoelectron energy window 32-42 eV. For this energy window, a partial wave analysis of the  $1s$ channel is presented in figure \ref{Fig:He1s-zoomed}. The states for which the Fano parameters are tabulated are labelled.

At wavelengths $\lambda_1$ and $\lambda_3$, the modulations in spectra can be attributed to {\color{black}the $(sp,2n+)$ and $(sp,2n-)$ series (linear combinations of $2snp$ and $2pns$ states) of doubly excited states} that can be excited upon absorption of one $\lambda_1$ photon or absorption of three $\lambda_3$ photons. In addition, at $\lambda_3$ wavelength, $2snf$ states appear. These two series of states corresponding to partial waves $L=1,\text{and }3$ are shown separately in figure \ref{Fig:He1s-zoomed}. The Fano parameterization (see equation (\ref{eq-Fano})) corresponding to the first four $(sp,2n+)$ states for the one and the three-photon processes is presented in tables \ref{tab:He20} and \ref{tab:He60} respectively. We also observe narrow lineshapes corresponding to $(sp,2n-)$ and $2pnd$ states.

Table \ref{tab:He20} presents a comparison of the obtained peak positions, widths and $q$ parameters with literature: experimental works \cite{Domke,Madden,Morgan} and theoretical works based on the method of complex rotation using correlated basis sets \cite{Jan1997,Lindorth1994}, R-matrix method \cite{PHamacher_1989}, many body perturbation theory \cite{Salomonson}, spline-Galerkin method\cite{Brage_1992} and Feshbach formalism\cite{Bhatia}. The peak positions among various theoretical studies deviate by about 100 meV. In our study, a deviation of 30 meV is expected due to our first ionization potential. Accounting for this deviation, our peak positions agree with Refs. \cite{Jan1997,Lindorth1994}. The linewidths and the $q$ parameters reported from various theoretical studies and experiments relatively vary by about 10\%. As expected for one-photon ionization process where a single channel is involved, the correlation coefficient $\rho^2\approx 1$. 

Table \ref{tab:He60} presents the peak positions and widths obtained with the three-photon ionization process. They agree with those obtained from one-photon process as they correspond to the same doubly excited states. {\color{black}The $q$ parameter, however, differs indicating its dependence on the nature of the excitation process. As it tends to a Lorentzian the explicit value is not presented.} The correlation coefficient also deviates from 1 as the process involves more than one channel which is seen from the partial wave analysis in figure \ref{Fig:He1s-zoomed}.

At $\lambda_2$ and $\lambda_4$ wavelengths, the system can be excited to $2sns$ and $2snd$ series of states through absorption of two and four photons respectively. In addition, at $\lambda_4$, $2sng$ states can be excited. The relevant partial wave channels are the $L=0,2\text{ and }4$ channels which are shown in figure \ref{Fig:He1s-zoomed}. Tables \ref{tab:He40} and \ref{tab:He80} present Fano parameters for the first two states in the series, namely the $^1S^e$ and $^1D^e$ states. We compare the peak positions and linewidths with experimental works \cite{Hicks_1975,F_Gelebart_1976} and  theoretical works based on the method of complex rotation \cite{Lindorth1994,Chen}, algebraic variational method\cite{Oza}, Feshbach formalism\cite{Sanchez_1995} and the multichannel quantum defect theory\cite{Wang}. Our peak positions and widths are in good agreement with the experimental works. The various theoretical studies differ relatively by about 10\%. In addition, we present the $q$ parameters and correlation coefficients. In Ref. \cite{Sanchez_1995}, a complex $q$ was reported for the two-photon process. However we find that a good fit can be obtained using a real $q$. This is in line with observation made in Ref. \cite{A_Cyr_1997} where a real $q$ parameter was sufficient to fit a three-photon resonance of Argon. For a given doubly excited state, the $q$ parameter associated with the two-photon and four-photon processes differ indicating its sensitivity to the excitation process. {\color{black} When the line shape tends to a Lorentzian, the $q$ parameter is not well defined and hence we do not present a specific value in the tables.}

\begin{table}
\fontsize{9}{10}\selectfont
\centering
\caption {Fano parameterization of $(sp,2n+)$ states following one-photon ionization of helium at 20 nm wavelength. Comparison with earlier theoretical and {\it experimental} works is presented. The states are labelled following convention in \cite{Madden}.}
\begin{tabular}{lllllll}
 
\hline
 State & Peak position (eV) & Width (eV) & q & $\rho^2$&Remarks  \\ \hline
(sp,22+)
&35.597&0.0406&	-2.63&0.999&{\bf haCC(1,9)} \\
&  35.626&	0.0373&	-2.77& &Jan M Rost et al. \cite{Jan1997} \\
&  35.636&	0.0374&	&& Lindroth \cite{Lindorth1994} \\
&  35.680&	0.0402&	-2.68& &P. Hamacher et al.\cite{PHamacher_1989} \\
&  35.674&	0.0403&	-2.63& &S. Salomonson et.al\cite{Salomonson}  \\
&  35.665&	0.0399&	& &T.Brage et al.\cite{Brage_1992} \\
&  35.635&	0.0363&	-2.84&& A.K.Bhatia et al.\cite{Bhatia} \\
&  35.637$(\pm 0.010)$&	0.0370$(\pm 0.001)$&	-2.75 $(\pm 0.001)$& &\textit{M. Domke et al.}\cite{Domke1996}\\
&  35.623$(\pm 0.015)$&	0.038$(\pm 0.004)$&-2.80$(\pm 0.25)$	&& \textit{Madden and Codling}\cite{Madden} \\
&  35.641$(\pm 0.010)$&	0.038$(\pm 0.004)$&	-2.55$(\pm 0.16)$&& \textit{Morgan and Ederer}\cite{Morgan} \\
\hline

(sp,23+)&39.094&0.00927	&-2.41
&0.999& {\bf haCC(1,9)} \\

&  39.134	&0.00819	&-2.58
& &Jan M Rost et al.\cite{Jan1997} \\ 
&  39.148&	0.008&	& &Lindroth \cite{Lindorth1994}\\
&  39.157&	0.00894&	-2.543&& P. Hamacher et al. \cite{PHamacher_1989}\\
&  39.160&	0.00896&	-2.43&& S. Salomonson et.al\cite{Salomonson}\\
&  39.154&	0.00878&	&& T.Brage et al.\cite{Brage_1992} \\
&  39.151&	0.009&	-2.60& &A.K.Bhatia et al.\cite{Bhatia} \\
&  39.148$(\pm 0.010)$&	0.010$(\pm 0.001)$&	-2.5$(\pm 0.1)$&& \textit{M. Domke et al.} \cite{Domke1996}\\
&  39.145$(\pm 0.007)$&	0.008$(\pm 0.004)$&	-2.0$(\pm 1.0)$& &\textit{Madden and Codling}\cite{Madden} \\
&  39.145$(\pm 0.010)$&	0.0083$(\pm 0.002)$&	-2.5$(\pm 0.5)$& &\textit{Morgan and Ederer}\cite{Morgan} \\
\hline

(sp,24+) &39.897&	0.00387	&-2.39
&0.996&{\bf haCC(1,9)} \\

&  39.942	&0.00349	&-2.55
&& Jan M Rost et al.\cite{Jan1997} \\
&  39.957	&0.00385&-2.534
&& Hamacher et al.\cite{PHamacher_1989}  \\
&  39.961&	0.00384&	-2.41&& S. Salomonson et.al\cite{Salomonson}\\
&  39.952$(\pm 0.007)$	&0.004$(\pm 0.0005)$	&
-2.4 $(\pm 0.1)$& &\textit{M. Domke et al.}\cite{Domke1996}
\\ \hline

(sp,25+) &40.246	&0.00199	&-2.38
&0.994&{\bf haCC(1,9)} \\

&  40.292	&0.00179	&-2.54
&& Jan M Rost et al.\cite{Jan1997} \\
&  40.303	&0.0012&-2.45
&& Hamacher et al.\cite{PHamacher_1989}  \\
&  40.308&	0.00197&	-2.4&& S. Salomonson et.al\cite{Salomonson}\\
&  40.303$(\pm 0.007)$&0.002$(\pm 0.0003)$	&-2.4 $(\pm 0.1)$
&& \textit{M. Domke et al.}\cite{Domke1996} \\
\hline
\end{tabular}
\label{tab:He20}
\end{table}

\begin{table}
\fontsize{9}{10}\selectfont
\centering
\caption {Fano parameterization of $(sp,2n+)$ states following three-photon ionization of helium at 60 nm wavelength. The states are labelled following convention in \cite{Madden}. The $q$ and $\rho^2$ are not tabulated since the peaks are nearly Lorentzian.} 
\begin{tabular}{lllllll}
\hline
 State & Peak position (eV) & Width (eV) & Remarks  \\ \hline
(sp,22+)
& 	35.598&0.0405	&{\bf haCC(1,9)} \\

(sp,23+)&39.094	&0.0093	&{\bf haCC(1,9)} \\

(sp,24+)  &39.897	&0.0039	&{\bf haCC(1,9)} \\

(sp,25+)  &40.246&0.002	&{\bf haCC(1,9)} \\

\hline
\end{tabular}
\label{tab:He60}
\end{table}

\begin{table}
\fontsize{9}{10}\selectfont
\centering
\caption {Fano parameterization of $^{1}S^{e}$ and $^{1}D^{e}$ states following two-photon ionization of helium at 40 nm wavelength. Comparison with earlier theoretical and {\it experimental} works is presented.}
\begin{tabular}{lllllll}
\hline
 State & & Peak position (eV) & Width (eV) & q &$\rho^2$& Remarks  \\ \hline
$^{1}S^{e}$& &33.253&0.131
&-12&0.13 &{\bf haCC(1,9)}\\
& & 33.340&0.138
&-6.97-i4.03 &&Sanchez et al.\cite{Sanchez_1995} \\
& & 33.330&0.124
& && Lindroth \cite{Lindorth1994} \\
& & 33.339&0.125
& && H.Oza\cite{Oza} \\
& & 33.329	&0.123
& && Ming-Keh Chen\cite{Chen} \\
& & 33.313&0.128
& && Y. Wang and C. H. Greene\cite{Wang} \\
& & 33.33$(\pm 0.04)$	&0.138$(\pm 0.015)$
& &&\textit{Hicks and Comer}\cite{Hicks_1975} \\
& & 33.27$(\pm 0.03)$	&0.138$(\pm 0.015)$
& &&\textit{Gelebart et al.} \cite{F_Gelebart_1976} \\
 \hline

$^{1}D^{e}$& &35.401&0.0689
&-23 &0.52&{\bf haCC(1,9)} \\
& & 35.43&0.0706
&-15.4-i7.40 && Sanchez et al.\cite{Sanchez_1995} \\
& & 35.39&0.064
& && Lindroth \cite{Lindorth1994} \\
& & 35.412&0.066
& && H.Oza\cite{Oza} \\
& & 35.398	&0.064
& && Ming-Keh Chen\cite{Chen} \\
& & 35.386&0.064
& && Y. Wang and C. H. Greene\cite{Wang}  \\
& & 35.400$(\pm 0.03)$	&0.072$(\pm 0.018)$
& &&\textit{Hicks and Comer}\cite{Hicks_1975} \\
& & 35.360$(\pm 0.03)$	&0.07$(\pm 0.01)$
& && \textit{Gelebart et al.} \cite{F_Gelebart_1976}\\
\hline

\end{tabular}

\label{tab:He40}
\end{table}
\begin{table}
\fontsize{9}{10}\selectfont
\centering
\caption {Fano parameterization of $^{1}S^{e}$ and $^{1}D^{e}$ states following four-photon ionization of helium at 80 nm wavelength. The $q$ and $\rho^2$ are not tabulated since the peaks are nearly Lorentzian.}
\begin{tabular}{lllllll}
\hline
 State & & Peak position (eV) & Width (eV) &Remarks  \\ \hline
$^{1}S^{e}$& &33.251&0.132
&{\bf haCC(1,9)} \\
$^{1}D^{e}$& &35.400&0.069
& {\bf haCC(1,9)} \\
\hline
\end{tabular}
\label{tab:He80}
\end{table}

\begin{figure}
    \centering
    \includegraphics[scale=0.3]{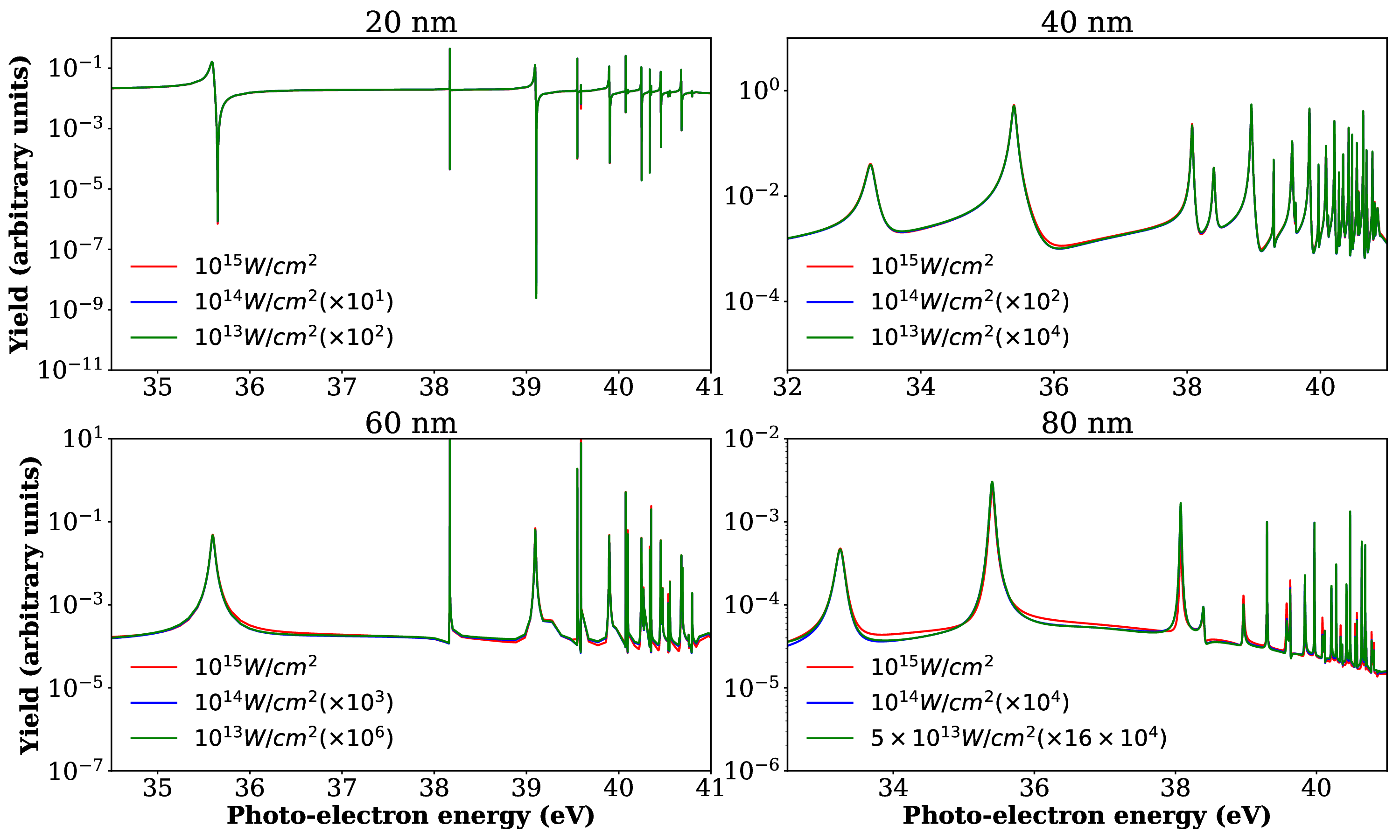}
    \caption{Intensity dependence for helium in the photoelectron energy window 32-42 eV. Each sub-plot title indicates the excitation wavelength. The yields at various intensities are scaled as indicated in the legend. The scaled results are nearly indistinguishable.}
    \label{Fig:He-intensity}
\end{figure}

{\color{black} In figure \ref{Fig:He-intensity}, we present an intensity dependence in the photoelectron energy window 32-42 eV by varying the peak intensities from $10^{13}-10^{15}$W/cm$^2$. We consider only cases where yields are above $10^{-10}$. We observe the expected power law dependence of yield on the intensity ($\text{Yield} \propto I^n$, where $n$ is the number of photons absorbed). The line profiles are independent of the intensity chosen.}

\newpage

\subsection{Neon}

Multi-photon ionization of neon is studied using haCC(1,4) basis set. The channels include the triply degenerate $1s^22s^22p^5(^2P^o)$ and the non-degenerate $1s^22s^12p^6(^2S^o)$ states. The CI wavefunctions for the ionic and neutral states were derived using the m-aug-cc-pvqz basis. In our calculations, the first ionization potential is 21.5 eV which is in good agreement with the value 21.564 eV reported in literature \cite{Saloman}. The laser pulse is chosen to be a 2-cycle pulse with a peak intensity of $10^{15}$W/cm$^2$ at central wavelengths 27 nm ($\lambda_5$) and 81 nm ($\lambda_6$). 

\begin{figure}
    \centering
    \includegraphics[scale=0.35]{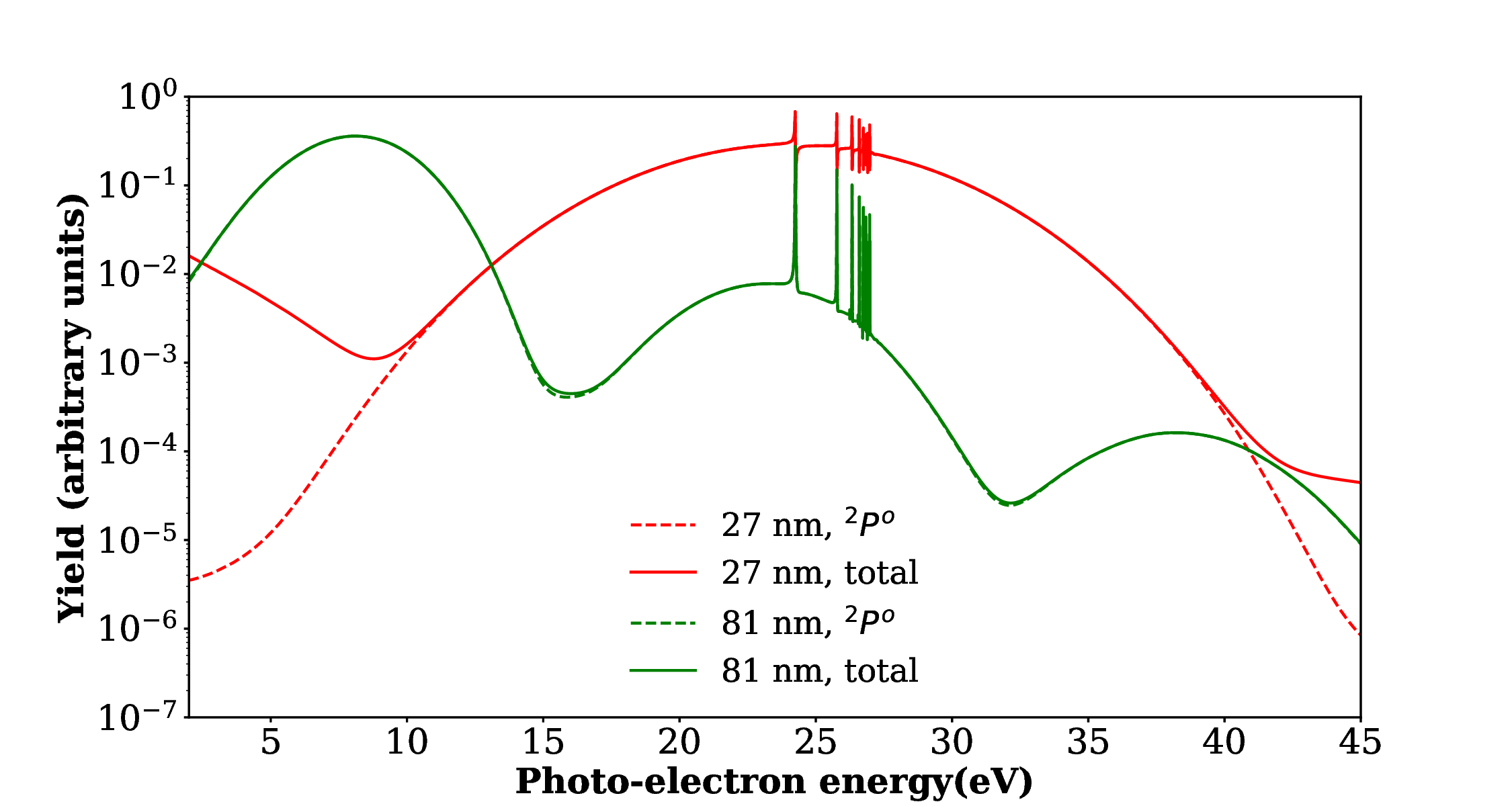}
    \caption{Total (solid) and $^2P^o$ (dashed) channel spectra for neon computed using haCC(1,4) basis with 2-cycle laser pulses having central wavelengths of 27 nm and 81 nm .}
    \label{Fig:Ne-total}
\end{figure}

\begin{figure}
    \centering
    \includegraphics[scale=0.35]{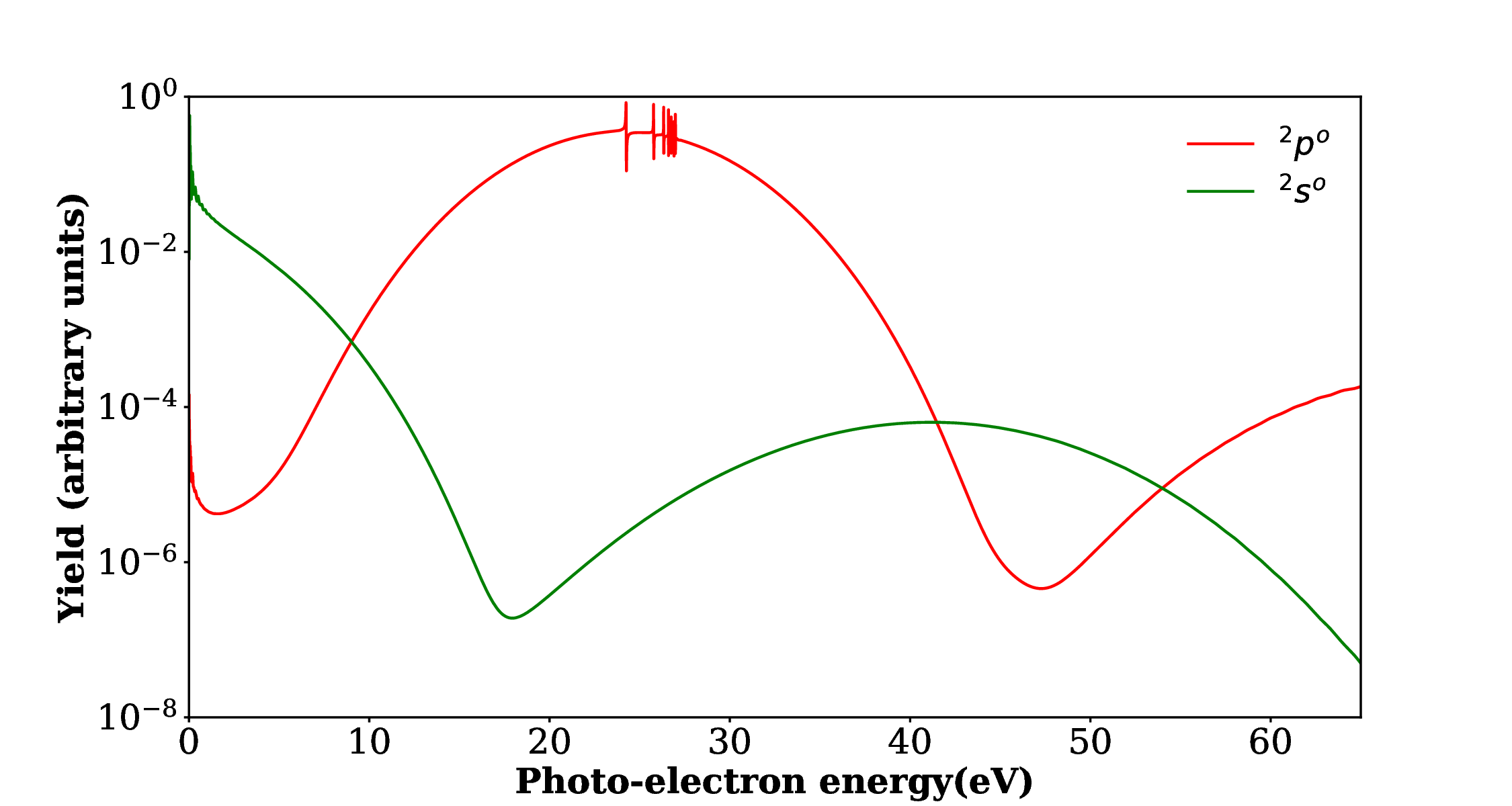}
    \caption{Channel resolved spectra for Neon computed using haCC(1,4) basis with 2-cycle laser pulses having central wavelengths of 27 nm. The channel label is indicated in the legend.}
    \label{Fig:Ne-channel27}
\end{figure}

\begin{figure}
    \centering
    \includegraphics[scale=0.25]{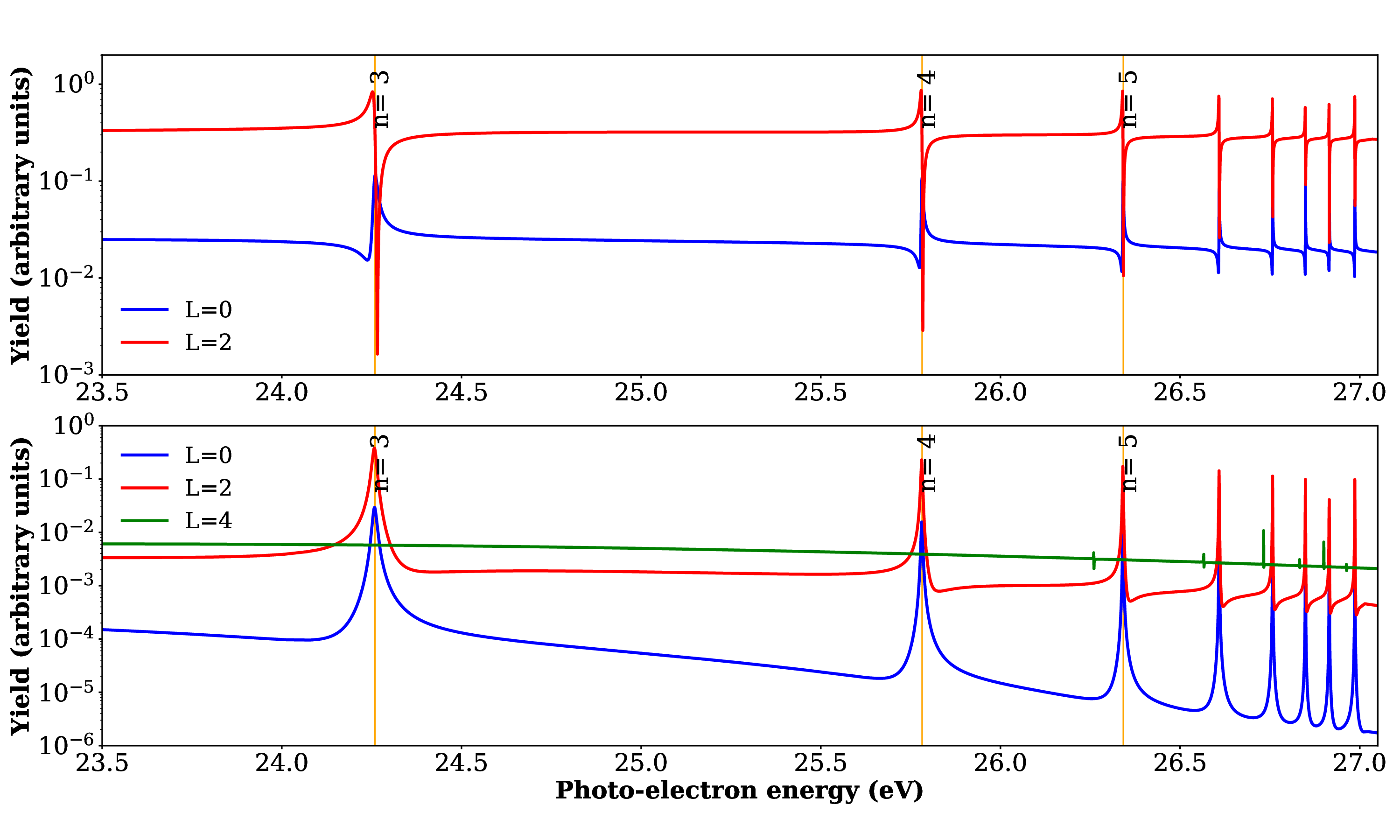}
    \caption{Partial wave analysis of $^2P^o$ channel spectra for neon in the photoelectron energy window 23-28 eV. The excitation wavelength is indicated in the title of the sub-plot. Dominant partial wave channels are presented.}
    \label{Fig:Ne2p-zoomed}
\end{figure}

Figure \ref{Fig:Ne-total} shows the total and the $^2P^o$ channel photoelectron spectra. We observe single as well as multi-photon peaks whose peak positions and the widths are related to the laser frequency, bandwidth and the number of photons absorbed. {\color{black} At 27 nm, the $^2S^o$ channel has a significant contribution at lower energies. The channel resolved spectra is shown in figure~\ref{Fig:Ne-channel27}. At 81nm, the $^2S^o$ channel's contribution is negligible and hence the total and the $^2P^o$ channel spectra coincide.}
Figure \ref{Fig:Ne2p-zoomed} presents a partial wave analysis of the $^2P^o$ channel in the photoelectron window 23-28 eV where several rapid modulations are observed. These features correspond to $2s 2p^6 np$ series of doubly excited states. These can be excited by absorption of one and three photons at $\lambda_5$ and $\lambda_6$ wavelengths respectively. The dominant partial waves for the one-photon process are $L=0,2$ while for the three-photon process are $L=0,2,4$. We parameterize the first three peaks using the Fano formula (\ref{eq-Fano}) and the fit parameters are presented in tables \ref{tab:Ne-27} and \ref{tab:Ne-81nm}. 

\begin{table}
\fontsize{9}{10}\selectfont
\centering
\caption {Fano parameterization of $2s 2p^6 np$ states following one-photon ionization of neon at 27 nm wavelength. Comparison with earlier theoretical and {\it experimental} works is presented.} 
\begin{tabular}{lllllll}
\hline
State& Peak position(eV) & Width(eV) & q &$\rho^2$& Remarks  \\ \hline
$2s2p^63p$ & 24.244&0.0130&-1.39&0.762&{\bf haCC(1,4)}\\ 
 
&24.021&0.0151&-1.34&0.77&C Marante et al.    (velocity)\cite{xchem}\\
&24.021&0.0150&-1.47&0.79&C Marante et al. (length)\cite{xchem}\\
&24.150&0.0114&& &Liang et al.\cite{LIANG2007599}\\
&24.843&0.0139&-3.69&0.514 &M Stener et al.\cite{Stener_1995}\\
&24.147&0.0186&-1.53&0.73 &B. Langer et al. (velocity)\cite{B_Langer_1997}\\
&24.147&0.0186&-1.59&0.72 &B. Langer et al. (length)\cite{B_Langer_1997}\\
&28.315&0.013&-1.4&0.77 &Nrisimhamurty et al.\cite{Nrisimhamurty}\\
&24.123&0.0349&& &Schulz et al.\cite{Schulz}\\
&24.136($\pm 0.008$)&0.013($\pm 0.002$)&-1.6($\pm 0.2$)&0.70$(\pm 0.07)$&\textit{K. Codling et al.} \cite{PhysRev.155.26}\\
&24.147&0.0132$(\pm 0.0010)$&-1.58$(\pm 0.1)$&0.75$(\pm 0.05)$&\textit{B. Langer et al.} \cite{B_Langer_1997}\\
\hline

$2s2p^64p$ & 25.767&0.00431&-1.50&0.77&{\bf haCC(1,4)}\\ 
 
&25.532&0.00430&-1.67&0.85&C Marante et al. (velocity)\cite{xchem}\\
&25.532&0.00430&-1.26&0.84&C Marante et al. (length)\cite{xchem}\\
&25.717&0.00528&& &Liang et al.\cite{LIANG2007599}\\
&25.987&0.00386&-3.95&0.505 &M Stener et al.\cite{Stener_1995}\\
&25.701&0.0043&-1.82&0.73 &B. Langer et al. (velocity)\cite{B_Langer_1997}\\
&25.701&0.0043&-1.88&0.72 &B. Langer et al. (length)\cite{B_Langer_1997}\\
&29.908&0.007&-1.35&0.63 &Nrisimhamurty et al.\cite{Nrisimhamurty}\\
&25.700&0.00665&& &Schulz et al.\cite{Schulz}\\
&25.711($\pm 0.005$)&0.045($\pm 0.0015$)&-1.6($\pm 0.3$)&0.70$(\pm 0.07)$&\textit{K. Codling et al.} \cite{PhysRev.155.26}\\
&25.701&0.057$(\pm 0.001)$&-1.47$(\pm 0.1)$&0.78$(\pm 0.11)$&\textit{B. Langer et al.} \cite{B_Langer_1997}\\
\hline

$2s2p^65p$ & 26.327&0.0019&-1.54&0.775&{\bf haCC(1,4)}\\ 

&26.092&0.0017&-1.78&0.86&C Marante et al. (velocity)\cite{xchem}\\
&26.092&0.0017&-1.35&0.86&C Marante et al. (length)\cite{xchem}\\
&26.287&0.00261&& &Liang et al.\cite{LIANG2007599}\\
&26.404&0.00162&-4.05&0.502 &M Stener et al.\cite{Stener_1995}\\
&26.277&0.0018&-1.87&0.75 &B. Langer et al. (velocity)\cite{B_Langer_1997}\\
&26.277&0.0018&-1.90&0.74 &B. Langer et al. (length)\cite{B_Langer_1997}\\
&30.484&0.003&-1.15&0.71 &Nrisimhamurty et al.\cite{Nrisimhamurty}\\
&26.281&0.00247&& &Schulz et al.\cite{Schulz}\\
&26.282($\pm 0.005$)&0.002($\pm 0.001$)&-1.6($\pm 0.5$)&0.70$(\pm 0.14)$&\textit{K. Codling et al.} \cite{PhysRev.155.26}\\
&26.277&0.0036$(\pm 0.0018)$&-1.46$(\pm 0.05)$&0.6$(\pm 0.2)$&\textit{B. Langer et al.} \cite{B_Langer_1997}\\
\hline
\end{tabular}
\label{tab:Ne-27}
\end{table}
\begin{table}
\fontsize{9}{10}\selectfont
\centering
\caption {Fano parameterization of $2s 2p^6 np$ states following three-photon ionization of neon at 81 nm wavelength. The $q$ and $\rho^2$ are not tabulated since the peaks are nearly Lorentzian.} 
\begin{tabular}{lllllll}
\hline
State& Peak position(eV) & Width(eV) &  Remarks  \\ \hline
$2s2p^63p$ & 24.244&0.0130&{\bf haCC(1,4)}\\ 

$2s2p^64p$ & 25.767&0.0043&{\bf haCC(1,4)}\\

$2s2p^65p$ & 26.327&0.0019&{\bf haCC(1,4)}\\ 
\hline
\end{tabular}
\label{tab:Ne-81nm}
\end{table}

For one-photon ionization process, we compare our results with experimental works \cite{PhysRev.155.26,B_Langer_1997} and theoretical works based on XCHEM\cite{xchem}, R-matrix method\cite{LIANG2007599,B_Langer_1997,Schulz}, density functional theory\cite{Stener_1995} and multichannel quantum defect theory\cite{Nrisimhamurty}. Our results are in good agreement with the experimental data with the experimental uncertainty included. Our results deviate from the most recent theoretical works of Carlos Marante et al. \cite{xchem} by about 15\%. They also report a gauge dependence on the same level indicating the accuracy of their results. Larger deviations are observed with the older theoretical reports which also exhibit a larger discrepancy with the experiments. For a given doubly excited state, the $q$ parameter corresponding to the three-photon process differs from the one-photon process. {\color{black} For the states considered here, the lineshapes become Lorentzian for the three-photon process.} These are presented in table \ref{tab:Ne-81nm}. {\color{black} As with the case of helium, the lineshapes are independent of the intensity chosen.}

\section{Discussion}
{\color{black}
The relative phase between (1) the direct ionization pathway and (2) the indirect pathway via the double excited state governs the $q$ parameter. These phases are defined by the corresponding transition matrix elements \cite{Fano1954}. In the computational studies presented above, we observe that with increasing the order of the multi-photon process the asymmetry decreases and line profiles tend towards Lorentzian. This indicates that as the order of the process is increased, path (2) corresponding to an exponential decay from the doubly excited state becomes more prominent. This is also supported by the fact that the ratio of the yield at the resonance to its background is higher for the higher order processes. For example, the ratio for the $(sp,22+)$ state of helium at $\lambda_1$ and $\lambda_3$ wavelengths is 6, 200 respectively and the ratio for $2s2p^63p$ state of neon at $\lambda_5$, $\lambda_6$ wavelengths is 2, 35 respectively.
Given that the lineshapes depend on the system specific transition matrix elements, the universality of these trends cannot be established here and would require additional work. However, the sensitivity of the Fano $q$ parameter to the order of the multi-photon process demonstrated here could be utilized for designing future spectroscopic techniques that aim to resolve photoionization dynamics involving multi-photon processes.
}

\section{Conclusion}

We presented an application of our newly developed \texttt{tRecX-haCC} package to study multi-photon Fano lines for helium and neon atoms. tSurff and iSurf techniques were used to efficiently compute the photoelectron spectra and the modulations arising from the slow decay of doubly excited states. We analyzed a few lineshapes in the photoelectron spectra and compared the linewidths and peak positions with the data available in the literature. We find a good agreement. In addition, we also analysze the Fano asymmetry parameter for the lineshapes on the multi-photon absorption peaks. Although arising from the decay of the same doubly excited state, the lineshapes characterized by the $q$ parameter are different depending on the number of photons absorbed.  Our study shows that \texttt{tRecX-haCC} is a valuable tool to study multiphoton processes and  opens up a new direction of study on multi-photon Fano lines with pulse light sources.

\section*{Acknowledgements}

VPM acknowledges financial support from Science and Engineering Research board (SERB) India (Project number: SRG/2019/001169). 

\printbibliography

\end{document}